\documentclass[preprint,longbibliography, articletitle, 12pt]{elsarticle}
\usepackage[colorlinks, citecolor=red]{hyperref}

\usepackage{amssymb}
\usepackage{amsmath}

\journal{Computational Materials Science}

\begin{document}

\begin{frontmatter}



\title{Magnetic phases and electron-phonon coupling in La$_3$Ni$_2$O$_7$ under pressure}


%
%
%
%
%
%
%
\author[lable1]{Cong Zhu}

\author[lable2,lable3]{Bin Li\corref{cor1}}
\ead{libin@njupt.edu.cn}
\cortext[cor1]{Corresponding author.}

\author[lable2]{Yuxiang Fan}

\author[lable1]{Chuanhui Yin}

\author[lable2,lable4]{Junjie Zhai}

\author[lable2,lable3]{Jie Cheng}

\author[lable2,lable3]{Shengli Liu\corref{cor1}}
\ead{liusl@njupt.edu.cn}

\author[lable5]{Zhixiang Shi}

\affiliation[lable1]{College of Electronic and Optical Engineering, Nanjing University of Posts and Telecommunications, Nanjing 210023, China}
\affiliation[lable2]{School of Science, Nanjing University of Posts and Telecommunications, Nanjing 210023, China}
\affiliation[lable3]{Jiangsu Provincial Engineering Research Center of Low Dimensional Physics and New Energy, Nanjing University of Posts and Telecommunications, Nanjing 210023, China}
\affiliation[lable4]{University of Chinese Academy of Sciences Nanjing (UCASNJ), Nanjing 211135, China}
\affiliation[lable5]{School of Physics, Southeast University, Nanjing 211189, China}

%

\begin{abstract}
Motivated by recent reports of pressure-induced superconductivity in bilayer nickelate La$_3$Ni$_2$O$_7$, we present a comprehensive investigation into the structural, electronic, magnetic, and phonon properties of this compound across a pressure range of 0 to 29.5 GPa. DFT+U calculations reveal that the A-type antiferromagnetic  ground state of La$_3$Ni$_2$O$_7$ persists throughout the studied pressure range. Electronic structure analysis shows that the Ni-$d_{xy}$ and Ni-$d_{z^2}$ orbitals dominate near the Fermi level in both the $Fmmm$ and $Amam$ phases of La$_3$Ni$_2$O$_7$. Phonon dispersion calculations for the $Fmmm$ phase reveal no imaginary modes from 12 to 29.5 GPa, confirming its dynamical stability in this pressure range. The vibrational frequencies of O atoms are substantially higher than those of Ni and La atoms, primarily due to the lower mass of oxygen. At 29.5 GPa, the electron-phonon coupling constant $\lambda$ for the $Fmmm$ phase is calculated to be 0.13. This small value suggests that conventional electron-phonon coupling is insufficient to explain the reported superconductivity in La$_3$Ni$_2$O$_7$, indicating a potentially unconventional mechanism. The study offers nuanced, actionable insights that can strategically inform and direct subsequent experimental investigations into the design and optimization of nickel-based superconducting materials.
\end{abstract}


%


\end{frontmatter}



\section{Introduction}
The observation of superconductivity in low valence layered nickelates has generated immense excitement in the community over the past few years\cite{daghofer2010three,li2019superconductivity,osada2020superconducting}. Recently, the newly discovered Ruddlesden-Popper bilayer perovskite nickelate La$_3$Ni$_2$O$_7$ has demonstrated a remarkable high superconducting transition temperature of $T_c$ $\sim$ 80 K at an applied pressure exceeding 14 GPa\cite{sun2023signatures,hou2023emergence,PhysRevB.108.L140505}. This breakthrough is poised to create a significant impact in the field of high-$T_c$ superconductivity, long after the discoveries of cuprate and iron-based superconductors. Given the identical 3\emph{d} electronic configuration of Ni$^{1+}$ in the parent phase, which is isoelectronic with Cu$^{2+}$, and the presence of comparable NiO$_2$ or CuO$_2$ layers, it was anticipated that the superconducting mechanism of the interlayer (IL) nickelate would closely mimic that of the cuprates. In the cuprates, superconductivity emerged upon hole doping, leading to expectations that a similar phenomenon could be observed in the IL nickelate as well\cite{Azuma1992}. However, numerous theoretical and experimental investigations have revealed fundamental differences between the individual infinite-layer nickelate and the cuprates\cite{AZZOUZ2019120954,OBEID201922,PhysRevLett.125.077003,PhysRevLett.124.207004,PhysRevB.101.041104}.

The discovery of a high-temperature superconducting nickel oxide with a Ni$^{2.5+}$ valence state has been groundbreaking, as its electronic structure and magnetism differ significantly from those of copper oxides. Experimental results demonstrate that La$_3$Ni$_2$O$_7$ undergoes a structural phase transition at approximately 12 GPa. Combined with thermodynamic enthalpy calculations of different structures, it has been determined that the space group of La$_3$Ni$_2$O$_7$ transitions from the low-pressure orthorhombic \emph{Amam} phase to the high-pressure \emph{Fmmm} phase\cite{VORONIN2001202,Liu_2022}. The Ni-apical oxygen atomic distance decreases from 2.297 Å in the ambient \emph{Amam} phase to 2.122 Å in the \emph{Fmmm} structure at 32.5 GPa. Additionally, the Ni-O-Ni angle between neighboring octahedra changes from $168^{\circ}$ in the atmospheric pressure \emph{Amam} space group to $180^{\circ}$ in the high-pressure \emph{Fmmm} space group\cite{PhysRevB.108.L140504,hou2023emergence}. Recent theoretical studies employing various techniques, including DFT+U, DFT+DMFT, GW+DMFT, and model Hamiltonians, have investigated the electronic structure of La$_3$Ni$_2$O$_7$ under pressure in relation to superconductivity. These studies emphasize the pivotal role of the Ni-\emph{d$_{z^{2}}$} states near the Fermi level in superconductivity. The key degrees of freedom near the Fermi level are the two Ni-{$e_{g}$} orbitals, with the Ni-\emph{d$_{z^{2}}$} orbital being nearly half-filled and the Ni-\emph{d$_{x^{2}-y^{2}}$} orbital being nearly quarter-filled.

The nickel-based superconductor La$_3$Ni$_2$O$_7$ has been the subject of several experimental and theoretical studies\cite{hou2023emergence,KUSHWAHA2021108495,10.1063/5.0046721,liu2023spmwave,PhysRevB.108.214522,https://doi.org/10.48550/arxiv.2401.00804,PhysRevB.109.165116,PhysRevB.109.045154,PhysRevB.109.045151,PhysRevLett.132.146002,PhysRevLett.132.126503,PhysRevMaterials.8.044801,Yang2024}. These investigations have explored various aspects of its electronic structure and superconducting properties. However, a systematic examination of the magnetic configurations in La$_3$Ni$_2$O$_7$ has not yet been conducted, leaving a gap in our understanding of this material's magnetic behavior. In this work, we investigate the magnetic configurations in La$_3$Ni$_2$O$_7$ for both its low-pressure phase (\emph{Amam}, space group No. 63) and high-pressure phase (\emph{Fmmm}, space group No. 69). We examine the pressure-induced evolution of the magnetic ground state. Electronic structure calculations are performed for both phases, with emphasis on the orbitals near the Fermi level. Furthermore, we analyze phonon properties, including a detailed study of phonon vibration modes.

\section{Methods}
Our calculations consist of three main components. First, we optimize the lattice parameters of La$_3$Ni$_2$O$_7$ under various pressures using the Quantum ESPRESSO package\cite{giannozzi2009quantum}. The initial lattice parameters are: $Fmmm$ ($a = 5.289$ Å, $b = 5.218$ Å, $c = 19.734$ Å) and $Amam$ ($a = 5.439$ Å, $b = 5.376$ Å, $c = 20.403$ Å)\cite{sun2023signatures,PhysRevLett.131.126001}. Second, we calculate the electronic structures incorporating correlation effects using density functional theory plus Hubbard $U$ (DFT+$U$)\cite{PhysRevResearch.5.013160,yu2020machine} to account for local Coulomb interactions in Ni ($3d$). Third, we compute the phonon spectra and electron-phonon coupling of the $Amam$ and $Fmmm$ phases at different pressures using density functional perturbation theory, analyzed with PHONOPY package\cite{togo2015first}. We set the charge density and wave function energy cutoffs to 60 Ry and 600 Ry, respectively. Perdew–Burke–Ernzerhof (PBE) generalized gradient approximation (GGA) pseudopotentials\cite{PhysRevLett.77.3865,kresse1999ultrasoft} are employed, selected from the standard solid-state pseudopotentials (SSSP)\cite{lejaeghere2016reproducibility,prandini2018precision}. For the self-consistent field calculations, we use $16 \times 16 \times 16$ $k$-point grids and $8 \times 8 \times 8$ $q$-point grids.


\section{Results and Discussion}
\begin{figure}[ht]
\begin{center}{\includegraphics[width=0.75\textwidth]{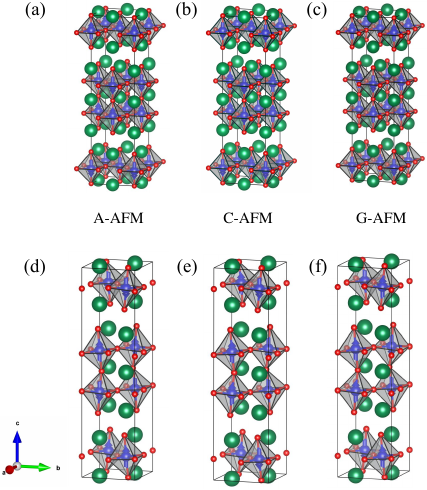}}
\caption{
\label{fig1}The geometrical structural models of La$_3$Ni$_2$O$_7$ with different magnetic configurations. (a)-(c) represent the $Fmmm$ structure, corresponding to the A-AFM, C-AFM, and G-AFM phases, respectively. (d)-(f) represent the $Amam$ structure, corresponding to the A-AFM, C-AFM, and G-AFM phases, respectively. Green, blue, and red spheres indicate La, Ni, and O atoms, respectively. Blue arrows on the Ni atoms indicate the directions of magnetic moments.
}
\end{center}
\end{figure}

To determine the ground state structure of La$_3$Ni$_2$O$_7$, we investigated two phases containing 8 Ni atoms, resulting in 70 possible antiferromagnetic (AFM) configurations. After excluding 35 configurations based on symmetry, we calculated the possible collinear spin order for the remaining 35 configurations. We also included calculations for the ferromagnetic (FM) configuration. Using experimentally known parameters to construct the crystal structure of La$_3$Ni$_2$O$_7$, we calculated the total energy for the non-magnetic (NM), FM, and 35 AFM configurations. As shown in Fig. \ref{fig1}(a)-(f), we focused primarily on three AFM structures of the $Fmmm$ and $Amam$ phases. In the G-AFM configuration, both intra- and interlayer couplings of the Ni layers exhibit AFM alignment. The C-AFM configuration displays intralayer AFM coupling and interlayer FM coupling in the Ni layers. Conversely, the A-AFM configuration exhibits intralayer FM coupling and interlayer AFM coupling in the Ni layers.

\begin{table}
\setlength{\tabcolsep}{1.5pt}
\renewcommand{\arraystretch}{0.8}

\begin{center}
\caption{\label{table1} 
Optimized lattice constants $a$, $b$, and $c$, relative enthalpies ($E$), and Ni magnetic moments ($M_\text{Ni}$) for $Fmmm$ and $Amam$ structures with various magnetic configurations at pressures from 0 to 29.5 GPa. Enthalpies are given relative to the $Amam$-A-AFM phase and were calculated using DFT+$U$ ($U = 3.5$ eV, $J_H = 0.7$ eV). Complete enthalpy data for all 35 magnetic configurations are provided in Appendix A of the Supplementary Materials.
}
\centering
\begin{tabular}{ccccccccc}
%

\hline
\hline
Structure &Pressure (GPa) & Phase & \emph{a} (\AA) & \emph{b} (\AA) & \emph{c} (\AA) &E (meV/atom)& $M_{\text{Ni}}$ ($\mu_{B}$)\\
\hline
& &A-AFM & 5.381 & 5.356 & 20.034 &22.16& 1.16 \\
& & G-AFM & 5.379 & 5.351& 20.032 & 30.70& 0.88 \\
&0 &C-AFM & 5.380 & 5.351 & 20.032 & 34.03 & 1.00 \\
& &FM & 5.376 &5.352& 20.042 & 35.49& 1.27 \\
& & NM & 5.390 & 5.359 & 20.064 & 55.70& 0 \\
 \cline{2-8} \\
\emph{Fmmm}& & A-AFM& 5.251 & 5.237 & 19.484 &-27.50 & 1.10 \\
& & G-AFM & 5.246 & 5.236 & 19.482 & -22.30 & 0.79 \\
& 20& C-AFM & 5.248 & 5.236 & 19.482 & -17.92& 0.91 \\
& &FM & 5.249 & 5.237 & 19.486 & -15.63 & 1.13 \\
& & NM & 5.255& 5.240 & 19.494 &-3.96 & 0 \\
 \cline{2-8} \\

& & A-AFM& 5.193 & 5.180 & 19.288 &-33.91& 1.06 \\
& & G-AFM & 5.189 & 5.180 & 19.278 &-29.95& 0.68 \\
&29.5 & C-AFM & 5.188 & 5.178 & 19.280 & -25.58&0.87 \\
& &FM & 5.192 & 5.180 & 19.288 & -23.50 & 1.08 \\
& & NM & 5.209& 5.198 & 19.306 &-11.83 & 0 \\
 \cline{1-8} \\

& &A-AFM & 5.395 & 5.419 & 20.085 & 0 & 1.21 \\
& & G-AFM & 5.395 & 5.420 & 20.105 & 7.50 & 1.15 \\
&0 &C-AFM & 5.398 & 5.422 & 20.121 & 12.08 & 1.07 \\
& &FM & 5.395 &5.418& 20.098 & 12.91 & 1.31 \\
& & NM & 5.435 & 5.516 & 20.141 & 47.50 & 0 \\
 \cline{2-8} \\

\emph{Amam}& & A-AFM& 5.240 & 5.245 & 19.520 & 0 & 1.11 \\
& & G-AFM & 5.241 & 5.247 & 19.521 & 6.66 & 0.84 \\
&20 & C-AFM & 5.245 & 5.249 & 19.529 & 9.79 & 0.93 \\
& &FM & 5.242 & 5.246 & 19.521 & 11.66 & 1.15 \\
& & NM & 5.257& 5.250 & 19.538 & 30.20 & 0 \\
 \cline{2-8} \\

& & A-AFM& 5.207 & 5.181 & 19.308 & 0 & 1.09 \\
& & G-AFM & 5.208 & 5.182 & 19.314 & 6.04 & 0.73 \\
&29.5 & C-AFM & 5.210 & 5.185 & 19.309 & 8.33 & 0.90 \\
& &FM & 5.208 & 5.183 & 19.314 & 11.48 & 1.10 \\
& & NM & 5.219& 5.218 & 19.360 & 21.87 & 0 \\
\hline
\hline
\end{tabular}
\end{center}
\end{table}

\begin{figure*}[!t]
	\begin{minipage}{0.75\linewidth}

		\centerline{\includegraphics[width=\textwidth]{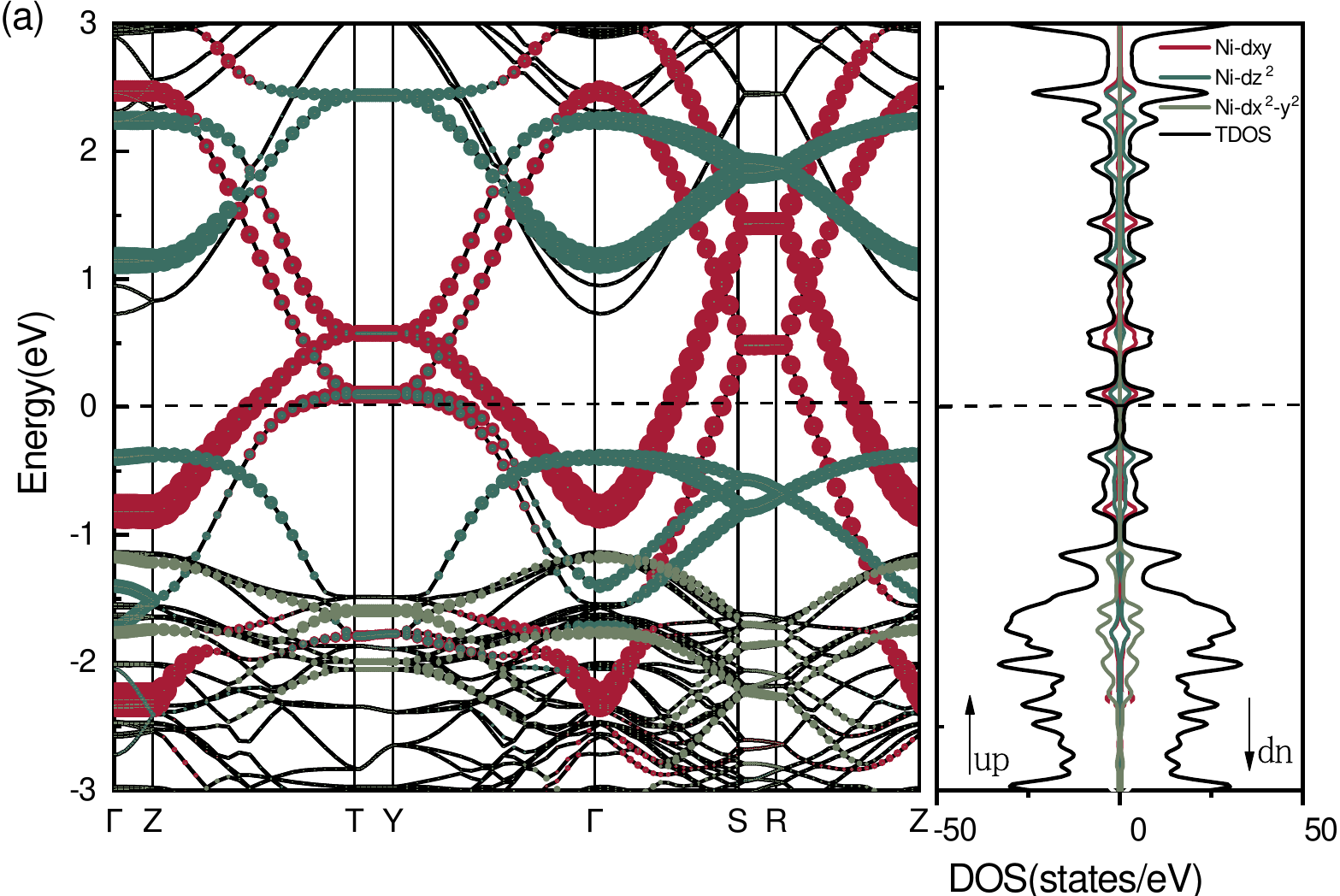}}
	\end{minipage}\hspace{-10mm}
	\begin{minipage}{0.75\linewidth}
	
		\centerline{\includegraphics[width=\textwidth]{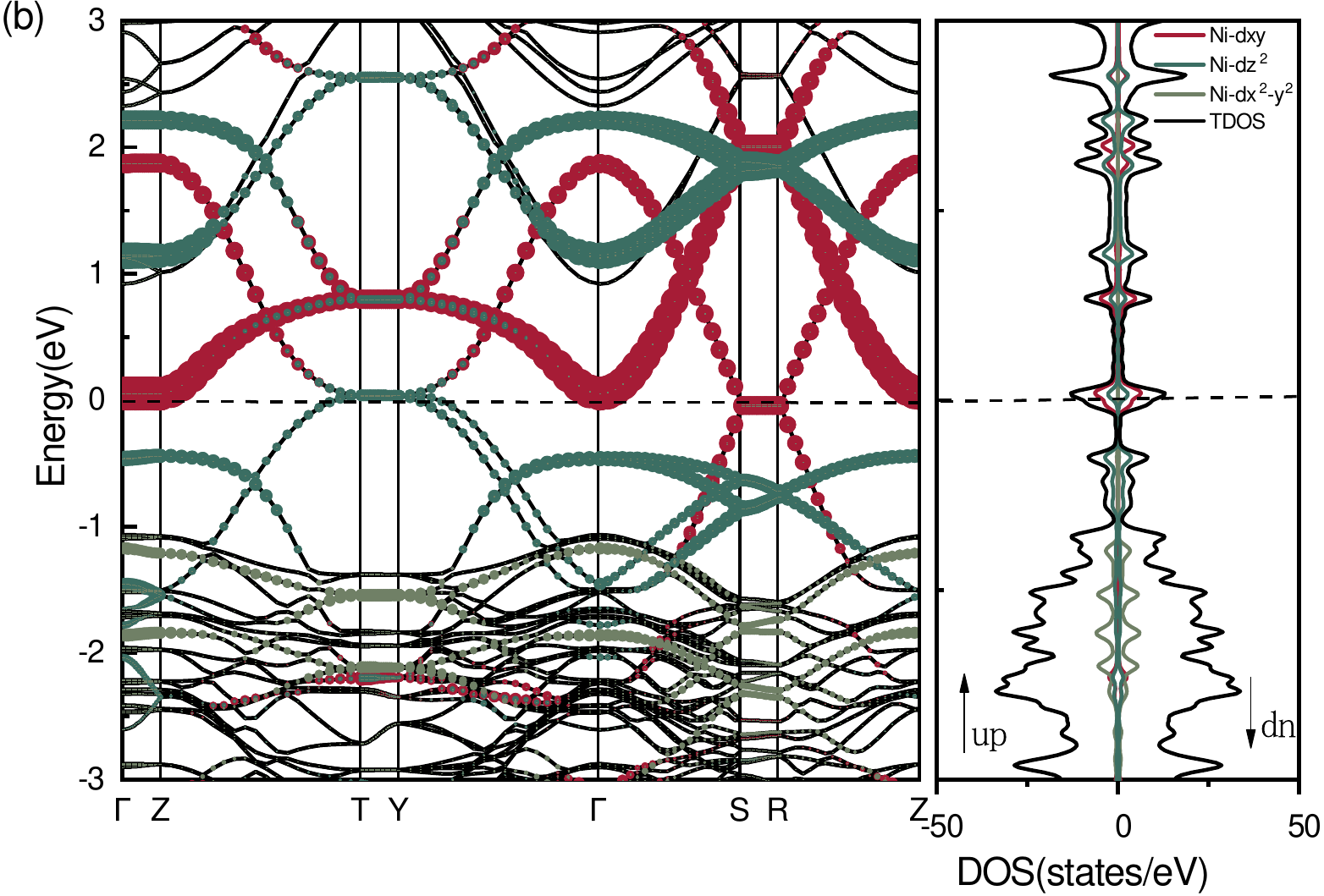}}
	\end{minipage}
\caption{\label{fig2}
  Electronic band structures (left panels) and total and Ni $d$-resolved density of states (right panels) for spin-up (down) in the A-AFM configuration of (a) $Fmmm$ and (b) $Amam$ structures under 29.5 GPa. The Fermi level $E_F$ is set to zero. Ni $d_{xy}$ (Ni $d_{z^2}$) is shown in red (dark green), and Ni $d_{x^2-y^2}$ is shown in brown. Band structures for the remaining pressure points are shown in Appendix B  of Supplementary materials.
}
\end{figure*}

We investigate the pressure-dependent phase space within the DFT+$U$ framework, elucidating the magnetic behavior of $Fmmm$ and $Amam$ structures. The magnetic ground state calculations are performed with $U = 3.5$ eV and $J_H = 0.7$ eV. Our results indicate that the A-AFM phase constitutes the ground state for both $Fmmm$ and $Amam$ structures across the entire pressure range of 0-29.5 GPa. Table \ref{table1} lists the calculated total energies, crystal lattice parameters, and local magnetic moments for different possible magnetic configurations of $Fmmm$ and $Amam$ structures under various pressures. Before describing the pressure effect on the ground state of these structures, we briefly consider their ground states under ambient pressure. As shown in Table \ref{table1}, without external pressure, the A-AFM configuration has the lowest total energy among all candidate configurations studied, suggesting it is the most stable magnetic structure. Notably, the energy differences between the various AFM (or FM) configurations and the NM configuration are exceptionally large in both $Fmmm$ and $Amam$ structures, indicating that the NM configuration is unlikely to be structurally stable. It should be noted that the lowest energy of A-AFM is determined from calculations of 35 antiferromagnetic configurations (See Appendix A of Supplementary materials).

We analyze the computational results of pressure-dependent calculations for key physical properties of the $Fmmm$ and $Amam$ structures.  Specifically, we compute the total energies, crystal lattice parameters, and local magnetic moments for various possible magnetic configurations within a pressure range of 0-29.5 GPa. For simplicity, we select 0, 20, and 29.5 GPa as representative pressure points. As shown in Table \ref{table1}, for all proposed states, as the pressure increases from 0 to 29.5 GPa, the A-AFM magnetic structure consistently presents the lowest energy, indicating it is the most stable magnetic configuration for both $Fmmm$ and $Amam$ structures within this pressure range. The optimized lattice parameters of the A-AFM phase for $Fmmm$ decrease with increasing pressure. At 0 GPa, they are $a = 5.381$ Å, $b = 5.356$ Å, and $c = 20.034$ Å, while at 29.5 GPa, they reduce to $a = 5.193$ Å, $b = 5.180$ Å, and $c = 19.288$ Å. Similarly, for the $Amam$ structure, the lattice parameters decrease from $a = 5.395$ Å, $b = 5.419$ Å, and $c = 20.085$ Å at 0 GPa to $a = 5.207$ Å, $b = 5.181$ Å, and $c = 19.308$ Å at 29.5 GPa. 
The calculated magnetic moments of different magnetic phases all decrease with increasing pressure, as shown in Table \ref{table1}.

To further understand the electronic properties of the stable A-AFM configurations of La$_3$Ni$_2$O$_7$, we calculated the orbital-projected band  structures, total density of states (TDOS), and orbital-resolved DOS of Ni $3d$ orbitals for $Fmmm$ and $Amam$ structures in the A-AFM phase at 29.5 GPa, as shown in Fig. \ref{fig2}(a) and (b). The Fermi level $E_F$ is set to zero. In the orbital-projected electronic structures, the Ni $d_{x^2-y^2}$ orbital contribution is represented by brown lines, while Ni $d_{xy}$ and Ni $d_{z^2}$ are shown in red and dark green, respectively.
For the $Fmmm$ structure, the TDOS at $E_F$ is primarily composed of Ni $d_{xy}$ and Ni $d_{z^2}$ states. The Ni $d_{xy}$ band crosses $E_F$, with its states mainly located between -2 eV and 2.5 eV. Ni $d_{x^2-y^2}$ states are concentrated between -3 eV and 0 eV, while Ni $d_{z^2}$ states range from -2.4 eV to 2.5 eV.
In the $Amam$ structure, Ni $d_{xy}$ orbitals contribute significantly to the TDOS at $E_F$. The main difference from $Fmmm$ is that the maximum energy of the Ni $d_{xy}$ band crossing $E_F$ at the $\Gamma$ point is shifted upward. Ni $d_{xy}$ states are primarily located between -0.2 eV and 2.1 eV, Ni $d_{x^2-y^2}$ states between -3 eV and 0.5 eV, and Ni $d_{z^2}$ states from -3 eV to 2.5 eV.
Additionally, based on the crystal structure of La$_3$Ni$_2$O$_7$-1313 at 300 K reported by Wang \cite{wang2023long}, we calculate the electronic structure of the non-magnetic state at 0 GPa (see Appendix B in Supplementary materials for band structure).

\begin{figure}[!t]
\vskip 4mm
\centerline{\includegraphics[width=0.750\textwidth]{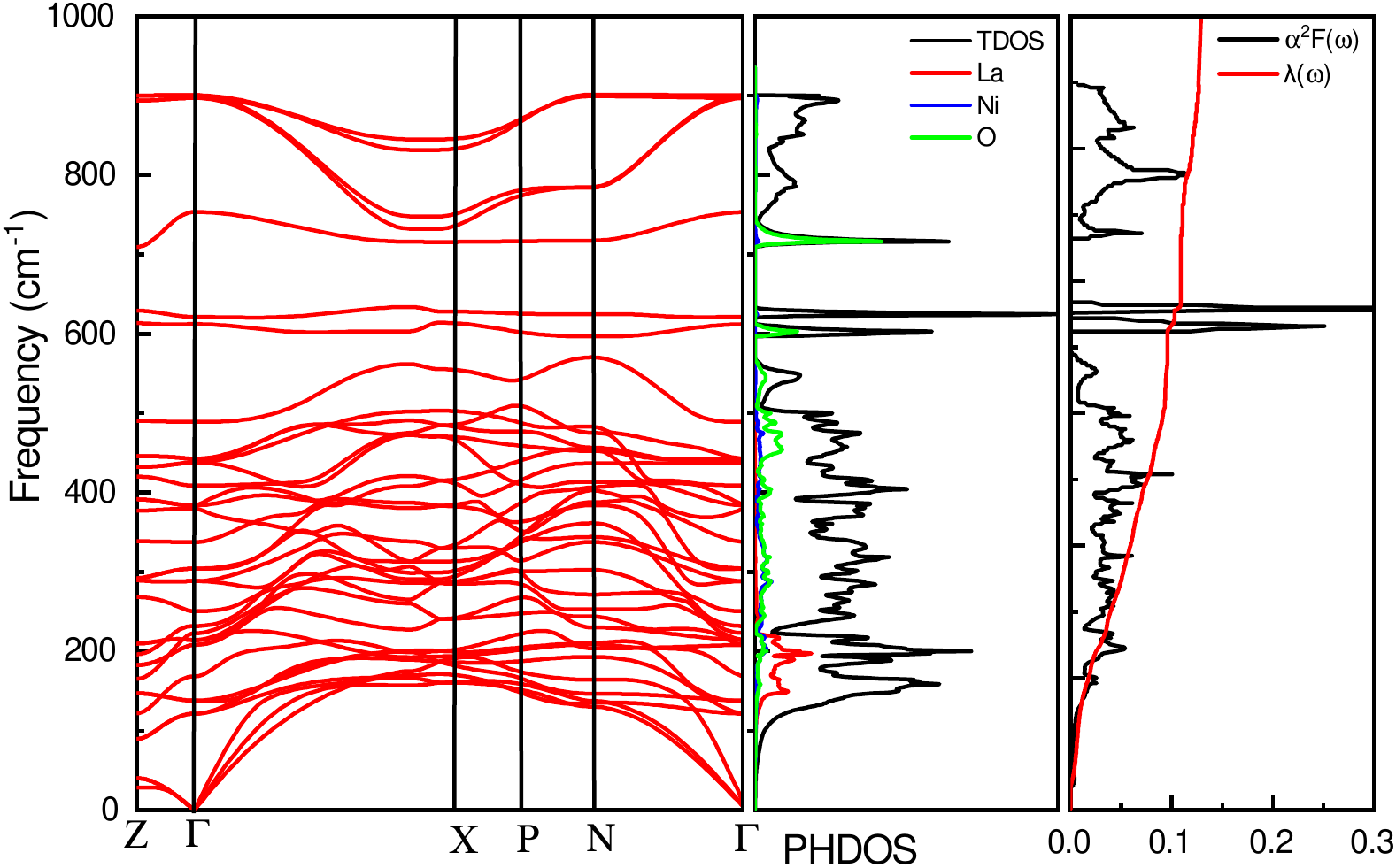}}
\caption{\label{fig3}
Left panel: phonon dispersion. Middle panel: phonon density of states. Right panel: Eliashberg spectral function $\alpha^2F(\omega)$ and electron-phonon coupling $\lambda(\omega)$ of the $Fmmm$ structure at 29.5 GPa. The PHDOS projections on La, Ni, and O are color-coded in red, blue, and green, respectively. Results for other pressure points are shown in Appendix C  of Supplementary materials.
}
\medskip
\end{figure}
We calculated the phonon properties of $Fmmm$ and $Amam$ structures using primitive cells. The $Fmmm$ phase shows no imaginary modes at 12-29.5 GPa, indicating its dynamical stability in this pressure range. Figure \ref{fig3} presents the phonon dispersion and phonon density of states (PHDOS) of $Fmmm$ at 29.5 GPa.
The absence of imaginary vibrations in the entire Brillouin zone confirms the structure's dynamical stability at this pressure. Phonon vibrations are mainly distributed in the middle and low frequencies. The PHDOS reveals that La atoms contribute most significantly to the density of states in the 100-200 cm$^{-1}$ range, followed by Ni and O atoms. In the 200-550 cm$^{-1}$ range, all atoms contribute, with O atoms showing the most pronounced effect, followed by Ni and La. A phonon energy gap appears in the 630-700 cm$^{-1}$ range, where the PHDOS of all three atoms approaches zero. In the 700-900 cm$^{-1}$ range, O atoms dominate the PHDOS, while La and Ni atoms contribute minimally.
For the $Amam$ phase, our calculations show stability at 0 GPa but imaginary modes at 15 GPa. Furthermore, using the La$_3$Ni$_2$O$_7$-1313 crystal structure\cite{wang2023long}, we observe that the $Cmmm$ phase exhibits imaginary modes in the pressure range of 20-30 GPa.

\begin{figure}[]

\centerline{\includegraphics[width=0.750\textwidth]{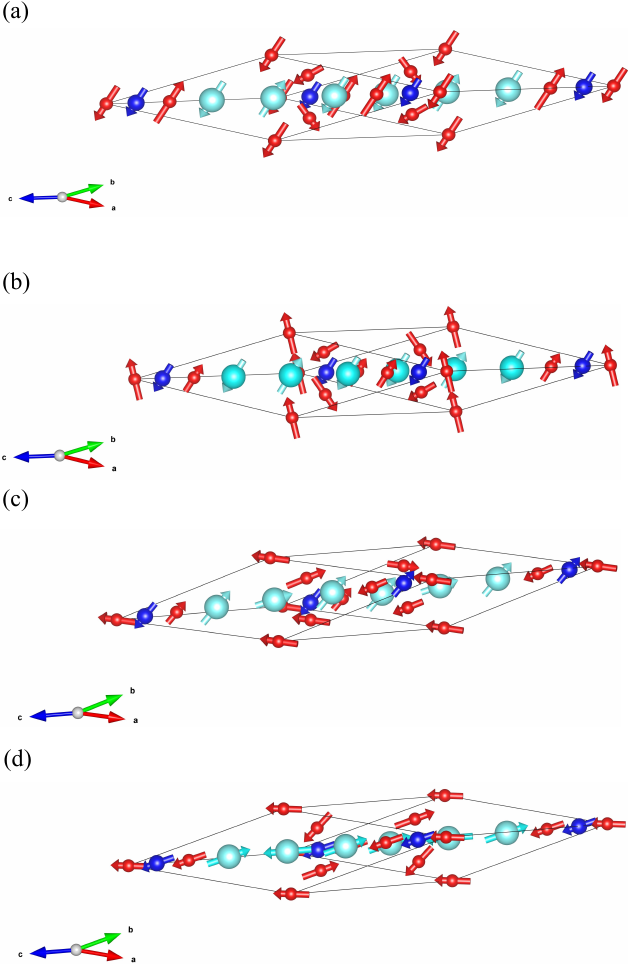}}
\caption{\label{fig4}
Visualization of atomic vibrations for selected phonon modes in La$_3$Ni$_2$O$_7$: (a) $A_{1g}$ $\sim$490.7 cm$^{-1}$, (b) $A_{2u}$ $\sim$613.5 cm$^{-1}$, (c) $A_{1g}$ $\sim$629.4 cm$^{-1}$, and (d) $A_{2u}$ $\sim$709.7 cm$^{-1}$. Red, cyan, and blue spheres represent O, La, and Ni atoms, respectively. Arrows indicate vibration directions, with arrow length proportional to vibration amplitude.
}

\end{figure}

The primitive cell of $Fmmm$ contains 12 atoms, resulting in 36 phonon bands: 3 acoustic and 33 optical branches. These can be irreducibly expressed as 4$A_{1g}$ $\oplus$ 6$A_{2u}$ $\oplus$ 14$E_u$ $\oplus$ 10$E_g$ $\oplus$ $B_{2u}$ $\oplus$ $B_{1g}$. $A_{2u}$ and $E_u$ are infrared-active modes, while $A_{1g}$, $B_{1g}$, and $E_g$ are Raman-active modes.
Figure \ref{fig4}(a)-(d) shows the atomic vibrations of selected Raman and infrared modes: $A_{1g}$ $\sim$490.7 cm$^{-1}$, $A_{2u}$ $\sim$613.5 cm$^{-1}$, $A_{1g}$ $\sim$629.4 cm$^{-1}$, and $A_{2u}$ $\sim$709.7 cm$^{-1}$, respectively. At these frequencies, oxygen atoms dominate the vibrations.
Combining the visualization of atomic vibrations with the PHDOS and analysis of Raman and infrared modes, we conclude that O atoms vibrate at higher frequencies than Ni and La atoms. This is primarily due to the significantly lower mass of O compared to Ni and La atoms.

In our study of the phonon properties of the $Fmmm$ structure, we calculate the electron-phonon coupling and present the integration of the Eliashberg function $\alpha^2F(\omega)$ and electron-phonon coupling $\lambda(\omega)$ in the rightmost panel of Fig. \ref{fig3}. By integrating $\alpha^2F(\omega)$, we obtain $\lambda = 2\int \alpha^2F(\omega)\omega^{-1}d\omega$ and the logarithmic average phonon frequency $\omega_\text{ln} = \exp[2\lambda^{-1}\int d\omega\alpha^2F(\omega)\omega^{-1}\log\omega]$.
Our calculations show that $\lambda$ remains small across all pressures studied. The $\lambda_{e\text{-}ph}$ values for the three pressure points we calculated are 0.1326, 0.1870, and 0.1311, respectively. These results indicate that La$_3$Ni$_2$O$_7$ is not an electron-phonon-dominated superconductor and suggests it may be an unconventional superconductor that deviates from standard BCS theory.

\section{Conclusion}
In this study, we have conducted a comprehensive investigation of the magnetic, electronic, and phonon properties of La$_3$Ni$_2$O$_7$ across the pressure range from 0 to 29.5 GPa. Our key findings include the determination of 35 antiferromagnetic configurations, with the A-AFM phase being identified as the ground state in the pressure range of 0-29.5 GPa. Analysis of the electronic structure of the ground state magnetic configuration reveals that the energy near the Fermi level $E_F$ is primarily contributed by the $d$ orbitals of Ni atoms, particularly the Ni $d_{xy}$ and Ni $d_{z^2}$ orbitals. Furthermore, the phonon dispersions for the $Fmmm$ phase of La$_3$Ni$_2$O$_7$ show no imaginary modes from 12 to 29.5 GPa, indicating dynamical stability in this pressure range, while the $Amam$ phase is found to be dynamically stable at 0 GPa. Notably, the calculated electron-phonon coupling constant $\lambda_{e\text{-}ph}$ is small across all pressures studied, suggesting that if La$_3$Ni$_2$O$_7$ exhibits superconductivity, it may involve mechanisms beyond the conventional BCS theory.

\section{Acknowledgments}
This work is supported by the National Natural Science Foundation of China (Grants No. 12374135, 12175107), Natural Science Foundation of Nanjing University of Posts and Telecommunications (Grants No. NY224165, NY219087, NY220038) and the Hua Li Talents Program of Nanjing University of Posts and Telecommunications. Some of the calculations were performed on the supercomputer in the Big Data Computing Center (BDCC) of Southeast University.\\
\appendix 
\section{Supplementary data}
See the supplementary material for the energy differences in various magnetic configurations, band structures, and phonon spectra of La$_3$Ni$_2$O$_7$ under pressures.










\end{document}